\documentclass[aps,prl,twocolumn,showpacs,floatfix]{revtex4}

\usepackage[ansinew]{inputenc}
\usepackage{epsfig}
\usepackage{graphicx}
\usepackage{color}
\usepackage{bm}
\usepackage{multirow}
\usepackage{mathrsfs}
\usepackage{revsymb}
\usepackage{amssymb}
\usepackage{amsmath}
\usepackage{slashed}
\usepackage{longtable}

\begin{document}

\title{High-precision metrology of highly charged ions via relativistic resonance fluorescence}

\author{O.~Postavaru}
\author{Z.~Harman} 
\email[]{harman@mpi-hd.mpg.de}
\author{C.~H.~Keitel}
\affiliation{Max Planck Institute for Nuclear Physics, Saupfercheckweg 1, 69117 Heidelberg, Germany}
\affiliation{ExtreMe Matter Institute EMMI, Planckstrasse 1, 64291 Darmstadt, Germany}


\begin{abstract}

Resonance fluorescence of laser-driven highly charged ions is studied in the relativistic regime by 
solving the time-dependent master equation in a multi-level model. 
Our ab initio approach based on the Dirac equation allows for 
investigating highly relativistic ions, and, consequently, provides a sensitive means
to test correlated relativistic dynamics, bound-state quantum electrodynamic phenomena and
nuclear effects by applying coherent light with x-ray frequencies.
Atomic dipole or multipole moments may be determined to unprecedented accuracy by measuring the
interference-narrowed fluorescence spectrum.

\end{abstract}



\maketitle

High-precision laser spectroscopy has resulted in crucial advancements in our 
understanding of nature. In particular, optical laser spectroscopy (OS) is a versatile 
tool to investigate correlated relativistic quantum dynamics, the testing of fundamental
theories like quantum electrodynamics (QED)~\cite{Bei05short,Gum05short} or parity
non-conservation in atomic systems. The determination of atomic dipole or multipole moments via
lifetime measurements by means of, e.g., visible emission spectroscopy~\cite{LJC05short},
approaching the accuracy of one per thousand, sheds light on QED effects like the electron
anomalous magnetic moment. Isotope shifts (IS) in atomic spectra which has been providing valuable 
insight into the collective structure of nuclei: for example, recently, isotope shifts were determined
by collinear laser spectroscopy~\cite{Gei08short}.
Beyond purely nuclear effects, the interaction of the correlated motion of electrons and that of the nucleus 
can be studied in IS measurements: recently, relativistic effects on nuclear recoil 
have been measured in visible forbidden transitions of few-electron argon ions by a 
trapped-ion method~\cite{Sor06short}.

In the regime of heavy few-electron systems, however, the accuracy of optical spectroscopy can seldom 
be exploited due to the scarcity of low-frequency transitions. Therefore, one has to 
apply other techniques. Measuring x-ray emission lines of 
highly charged uranium ions confined in an electron beam ion trap
allowed testing high-field QED on the two-loop level~\cite{Bei05short}
and delivered a new  value for the radius of the radioactive isotope ${}^{235}$U~\cite{Ell96}.
Recently, a method based on the storage ring measurement of dielectronic recombination 
spectra yielded the change of charge radii for neodymium isotopes~\cite{Bra08short,Sch04}.

\color{black}{With the advent of modern short-wavelength laser systems,
the accuracy and versatility of laser spectroscopy may be combined with the increased sensitivity of
highly charged ions (HCI) to relativistic and QED effects, nuclear properties, and this may also give
new information on HCI structure and dynamics relevant in astrophysical and thermonuclear plasmas.
Brilliant x-ray light has already enabled to study transitions in the soft x-ray regime involving
HCI~\cite{Epp07short}.}\color{black}{}
Coherent light with photon energies over 10~keV becomes accessible in the near future~\cite{XFEL}, allowing
for an extension to heavier systems and the exploitation of coherence properties. This would also ask for
the validity of numerous quantum control schemes of resonance
fluorescence~\cite{Narducci1990,Zhu96,Zhou96,Pasp98,Scully98,Keitel99,ScZu,Ficek,Kiffnershort} 
for concrete systems in the relativistic regime.

In the present Letter, we investigate the possibility of measuring atomic transition dipole --
or multipole -- moments and transition energies via relativistic resonance fluorescence of a three-level
atomic configuration driven by two fields, namely, a short-wavelength laser and a 
long-wavelength light source in the optical regime. In such a three-level setting, the 
linewidth of the spontaneous transition of interest may be rendered much narrower than 
the natural linewidth, with the simultaneous increase of the total emitted intensity
by orders of magnitude. Due to this effect, the determination of
atomic multipole moments by means of the detection of the fluorescence spectrum
is anticipated to increase in accuracy by several orders of magnitude.

\color{black}{As relativistic effects on the electronic wave function increase rapidly with the 
nuclear charge number $Z$, we formulate a fully relativistic ab initio theory of coherent laser-atom interaction 
based on the Dirac equation.}\color{black}{}
An approach via the 
time-dependent numerical solution of this wave equation was recently 
employed to describe ionization phenomena~\cite{Henrik09}. For our purposes, one needs to 
go beyond this approach and incorporate radiative relaxation in bound-bound transitions.

The interaction of a radiation field, having the four-vector potential $A_{\mu}$ ($\mu 
\in \{0,1,2,3\}$), with a single-electron atom can be described by the Hamiltonian (in 
relativistic units) $\mathscr{H}=\mathscr{H}_A+\mathscr{H}_F+e\gamma^{\mu}A_{\mu}$. Here, 
$\mathscr{H}_A$ and $\mathscr{H}_F$ are the energy operators of the atom and the 
radiation field, respectively, $e$ is the elementary charge, and $\gamma^{\mu}$ is the 
four-vector of Dirac matrices, represented in terms of $2\times2$ Pauli matrices. The 
atom is described in a fully relativistic manner, i.e. 
$\mathscr{H}_A=i\gamma^{\mu}\partial_{\mu}+V(r)$, where $V(r)$ is the nuclear potential, 
$r$ the distance from the origin, and $\partial_{\mu}$ is the space-time gradient 
operator. The Hamiltonian of the electromagnetic field $\mathscr{H}_F$ is given in 
terms of the photon creation and annihilation operators by 
$\mathscr{H}_F=\sum_{\mathbf{k}}\omega_{k}\left(a^{\dagger}_{\mathbf{k}}a_{\mathbf{k}}+\frac{1}{2}\right)$. 
To put the total Hamiltonian in a more convenient form, $\mathscr{H}_A$ and 
$\gamma^{\mu}A_{\mu}$ can be expressed in terms of the atomic operators 
$\sigma_{ab}=|a\rangle\langle b|$ describing transitions between a state $|a\rangle$ and 
$|b\rangle$ (${a,b} \in \mathbb{N}$).
\color{black}{Let us introduce the coupling constants (Rabi frequencies)
$g_{ab}=-\boldsymbol{\mu}_{ab}\cdot\mathbf{\epsilon}_{\mathbf{k}} \mathscr 
E_{\mathbf{k}}$, proportional to the corresponding laser field strength,
with the relativistic multipole transition matrix element 
$\boldsymbol{\mu}_{ab}=\tfrac{e}{\omega}\langle 
a|\boldsymbol{\alpha}\cdot\mathbf{\epsilon}_{\mathbf{k}} e^{i\mathbf{k}\cdot\mathbf{r}}|b\rangle$.}
\color{black}{}
Here, $\mathbf{\epsilon}_{\mathbf{k}}$ denotes the laser polarization, $\omega=|\mathbf{k}|$ and
$\mathscr E_{\mathbf{k}}$ is the amplitude of the electric field. We evaluate 
the multipole transition matrix elements by using the multipole expansion of the
vector potential
\begin{eqnarray}
\mathbf{\epsilon}_{\mathbf{k}}e^{i\mathbf{k}\cdot\mathbf{r}}=4\pi\sum_{JM\lambda}
i^{J-\lambda}(\mathbf{Y}_{JM}^{(\lambda)}(\hat{k})\cdot\mathbf{\epsilon}_{\mathbf{k}})
\mathbf{a}_{JM}^{(\lambda)}(\mathbf{r}) \,.
\end{eqnarray}
The electric ($\lambda=1$) and magnetic ($\lambda=0$) multipole 
potentials $\mathbf{a}_{JM}^{(\lambda)}(\mathbf{r})$ are given by
\begin{eqnarray}
\mathbf{a}_{JM}^{(0)}(\mathbf{r})&=&j_{L}(kr)\mathbf{Y}_{JLM}(\hat{r}),\\\nonumber
\mathbf{a}_{JM}^{(1)}(\mathbf{r})&=&\sqrt{\frac{J+1}{2J+1}}j_{J-1}(kr)\mathbf{Y}_{JJ-1M}(\hat{r})\\\nonumber
&-&\sqrt{\frac{J}{2J+1}}j_{J+1}(kr)\mathbf{Y}_{JJ+1M}(\hat{r})\,
\end{eqnarray}
in the transverse gauge, which can be regarded as a relativistic generalization of the velocity form.
The symbols $\mathbf{Y}_{JM}^{(\lambda)}(\hat{r})$ and $\mathbf{Y}_{JLM}(\hat{r})$ stand for the vector
spherical harmonics. We adopt the generalized length gauge, also referred to as the Babushkin gauge~\cite{Grant}.
This general description also accounts for higher-order (non-dipole) transitions resulting in narrow spectral features.
\color{black}{The Hamiltonian accounting for the interaction of the three-level system with two classical fields
is~\cite{Narducci1990}
\begin{eqnarray} 
\mathscr{H}_I&=&
g_{31}(e^{i\Delta_{x}t}\sigma_{31}+e^{-i\Delta_{x}t}\sigma_{13})\\
&+&g_{21}(e^{i\Delta_{o}t}\sigma_{21}+e^{-i\Delta_{o}t}\sigma_{12})\,, \nonumber
\end{eqnarray}
where $\Delta_x=\omega_{31}-\omega_x$ and $\Delta_o=\omega_{21}-\omega_o$ are the detunings of the 
x-ray and optical laser frequencies $\omega_{x}$, $\omega_{o}$ from the atomic transition energies.
The master equation describing the dynamics of the system with the Hamiltonian in the interaction picture
is given by $\dot{\rho}=-i[\mathscr H_I,\rho]+\Lambda\rho$,
where $\Lambda$ is the Lindblad operator accounting for the relaxation channels~\cite{ScZu,Narducci1990,Ficek,Kiffnershort}.
We have also included the dephasing rate $\gamma_D$ of x-ray lasers by adding it to the decay rate 
$\gamma_{13}=\frac{1}{2}(\Gamma_{31}+\Gamma_{32})$ of the 3$\leftrightarrow$1 x-ray transition~\cite{ScZu,Agarwal},
since the coherence time of free electron lasers (FELs) is often shorter than their pulse length.}
\color{black}{}$\Gamma_{ab}$ denotes the partial radiative rate for the a$\to$b transition.

Our goal is to evaluate the power spectrum of radiation scattered by a few-level 
atom driven by an incident field of arbitrary strength. To this end we introduce the 
field operators $\mathbf{E}^{(\pm)}(\mathbf{r},t)$ for a given time $t$ and observation 
point $\mathbf{r}$ in the Weisskopf-Wigner approximation
\begin{eqnarray}\label{PozFieldOp}
\mathbf{E}^{(+)}(\mathbf{r},t)=\frac{\omega^2\sin\eta}{4\pi r}
\hat{e}_xP^{(+)}\left(t - r \right) \,,
\end{eqnarray} 
and the corresponding conjugate expression for $\mathbf{E}^{(-)}(\mathbf{r},t)$. Here, 
$P(\tau)^{(+)}=\boldsymbol{\mu}_{12}(\sigma_{12}+\sigma_{21})+\boldsymbol{\mu}_{13}(\sigma_{13}+\sigma_{31})$. 
The atomic dipole (or, more generally, multipole) is assumed to be in the plane defined 
by the Cartesian coordinates $x$ and $z$, and $\eta$ is the angle between the dipole and 
the $z$-axis.

\begin{figure}[t!]
\includegraphics[width = 0.80 \columnwidth]{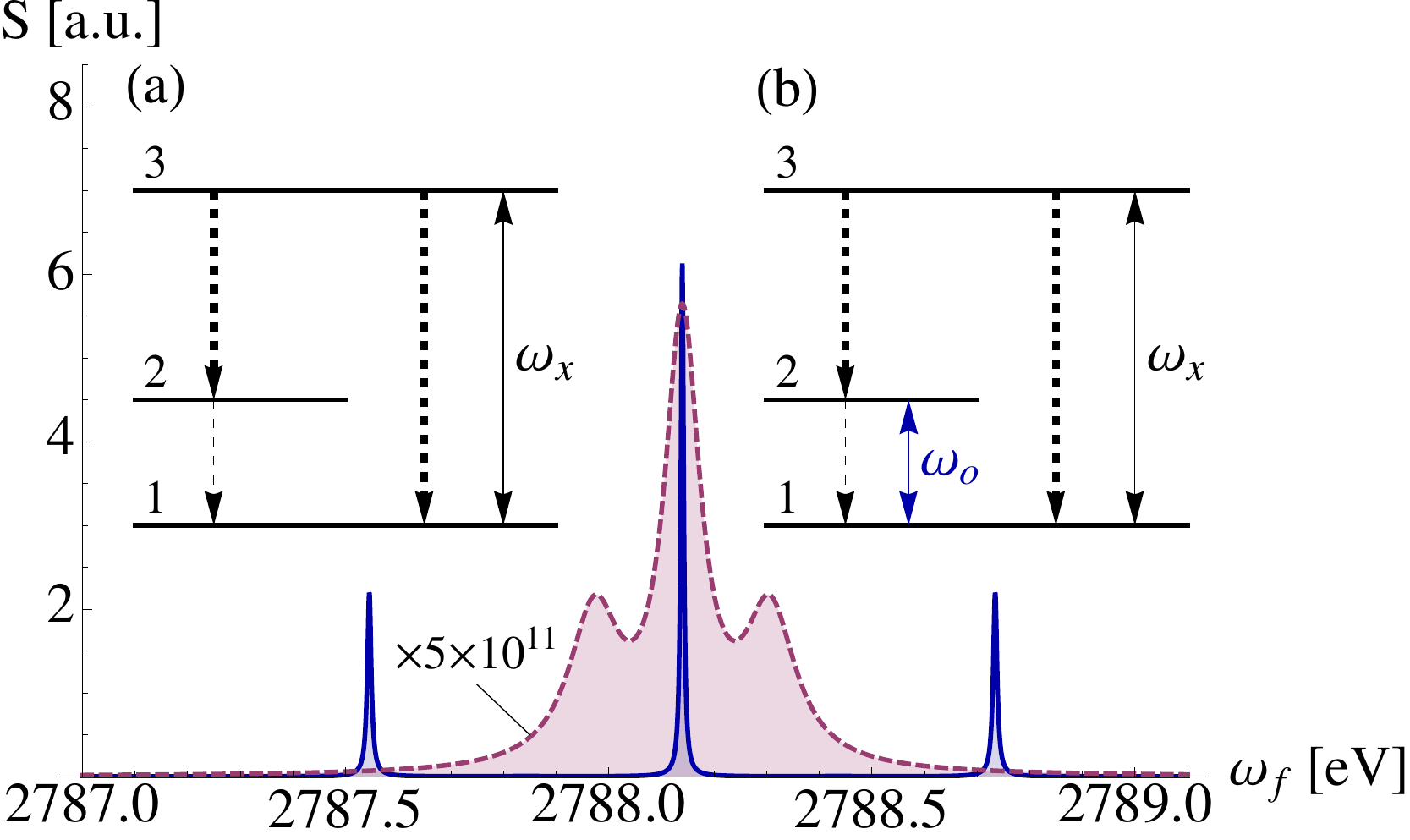}
\vspace{-2mm}
\caption{\label{fig:narrowing}
(Color online.) Fluorescence photon spectrum
for the $2s$$\leftrightarrow$$2p_{3/2}$ transition in
Li-like ${}^{209}$Bi as a function of the fluorescence photon frequency $\omega_f$.
(a)~Dashed (red) curve:
An x-ray laser ($I_x=5\times10^{11}$~W/cm${}^{2}$) is in resonance with the
ionic electric dipole (E1) transition at $\omega_{31}=$ 2788.1 eV between the
hyperfine-split ground state 1 ($2s$ with $F=4$, $M_F=4$) and the uppermost state 3 ($2p_{3/2}$ with
$F=5$, $M_F=5$). This curve is multiplied by a factor of $5 \times 10^{11}$. Thick (thin) dashed arrows represent
fast E1 x-ray (slow M1 optical) decays.
(b)~Continuous (blue) curve: an additional optical driving ($I_o=10^{14}$~W/cm${}^{2}$) is
applied on the $\omega_{21}=$ 0.797~eV~\cite{Sha98BiHFS} M1
transition between the hyperfine-split magnetic sublevels 1 ($F=4$, $M_F=4$) and 2 ($F=5$, $M_F=5$).
The inner sidebands are suppressed. See text for more details.
}
\end{figure}

We are interested in the field emitted by the atom fixed in position along the $x$-axis. 
The field operator in Eq.~(\ref{PozFieldOp}) can therefore be treated as scalar. The power 
spectrum $S(\mathbf r,\omega_0)$ of the fluorescent light at some suitably chosen point 
$\mathbf r$ in the far field is obtained by taking the Fourier transform of the 
normally-ordered field correlation function $\langle \mathbf{E}^{(-)}(\mathbf r,t) 
\mathbf{E}^{(+)}(\mathbf r,t+\tau)\rangle$ with respect to the time delay $\tau$. 
According to the Wiener-Khintchine theorem, the power spectrum $S(\omega_0)$ is
$
\label{spectrum}
S(\omega_0)=\frac{1}{\pi}\mathrm{Re}\int_0^{\infty}d\tau
\langle \mathbf{E}^{(-)}(\mathbf r,t)
\mathbf{E}^{(+)}(\mathbf r,t+\tau)\rangle e^{i\omega_0\tau} \,.
$

We calculate the power spectrum and the transition energies and matrix elements involved 
by relativistic methods. The Dirac wave functions corresponding to a 
rotationally symmetric nuclear potential are given in spherical coordinates in the form 
(see, e.g.~\cite{Grant})
\begin{equation}\label{eq:Diracwf}
\Phi_a (\bm{r}) = \langle\bm{r}  | a  \rangle =
\left(
\begin{array}{c}
 G_{n_a\kappa_a}(r)  \Omega_{\kappa_a  m_a}\left(\hat{r}\right)\\
 i F_{n_a\kappa_a}(r)\Omega_{-\kappa_a m_a}\left(\hat{r}\right)
\end{array}
\right) \,.
\end{equation}
The subscript $a$ stands collectively for the principal quantum number $n_a$ as well as 
the angular momentum quantum numbers $j_a$ and $l_a$ and the magnetic quantum number 
$m_a$ of a given state $a \equiv (n_a, j_a,l_a, m_a)$. For Coulomb potentials, the bound 
radial functions $G_a(r)$ and $F_a(r)$ are known analytically, involving confluent 
hypergeometric functions with negative-integer first argument. The spin-angular part of 
the Dirac wave function is defined by the spherical spinors $\Omega_{\kappa_a 
m_a}\left(\hat{r}\right)$~\cite{Grant}. \color{black}{In the transition energies $\omega_{ab}$,
also the one-loop QED corrections of self-energy and vacuum polarization~\cite{MPS}
have been included.}\color{black}{}

In Fig.~\ref{fig:narrowing} (a) we plot the power spectrum of resonance fluorescence for the case 
of the $2s$$\leftrightarrow$$2p_{3/2}$ electric dipole transition in Li-like ${}^{209}$Bi ($Z$=83) 
ions. The dynamic (AC) Stark shift leads to a splitting of the central peak, giving rise 
to a Mollow spectrum.
Due to the long lifetime of the upper level of the hyperfine-split ground state, level 2, almost 100 \% of the
population is trapped in this level if only the 3$\leftrightarrow$1 transition is driven coherently
with an x-ray laser. The calculation yields for the population of the uppermost state a value of approx.
$\Gamma_{21}/(\Gamma_{32}+2\Gamma_{21})\approx10^{-12} \ll 1$, resulting in a negligibly small total x-ray fluorescence~\cite{remark}.
This undesirable effect may be reversed if additionally the 2$\leftrightarrow$1 optical transition is coherently driven
(see Fig.~\ref{fig:narrowing}), leading to an efficient re-population of level 3.

Furthermore, the spectral lines become substantially narrower due to coherence and interference effects
(see~\cite{Narducci1990} for the pioneering non-specific treatment).
The width of the central peak and the outer sidebands are given
by $\Gamma_C=(\Gamma_{31}+\Gamma_{32}+\gamma_D)R+\Gamma_{21}(1-R)$ and
$\Gamma_{SB}=| \frac{3}{2}(\Gamma_{31}-\frac{1}{3}\gamma_D)R + \frac{1}{2}\Gamma_{32}(R+R^2) + \frac{3}{2}\Gamma_{21}(1-R) |$,
respectively, with the ratio $R$ being $g_{31}^2/(g_{31}^2+g_{21}^2)$~\cite{remark}.
This effect is shown in Fig.~\ref{fig:narrowing} (b). Further increasing the intensity of the 
long-wavelength driving field and thus $g_{21}$ could even assign the narrow linewidth of 
$7.7\cdot10^{-15}$ eV of the M1 hyperfine transition to the E1 x-ray transition of interest.
\color{black}{The above line width formulas also imply that the dephasing width $\gamma_D$ -- typically on the order of 1 eV
for XFELs~\cite{XFEL} -- does not hamper the observation of sub-natural linewidths in the x-ray regime as its contribution
is decreased with the same factor~$R$.}\color{black}{}

\begin{figure}[t!]
\includegraphics[width = 0.8 \columnwidth]{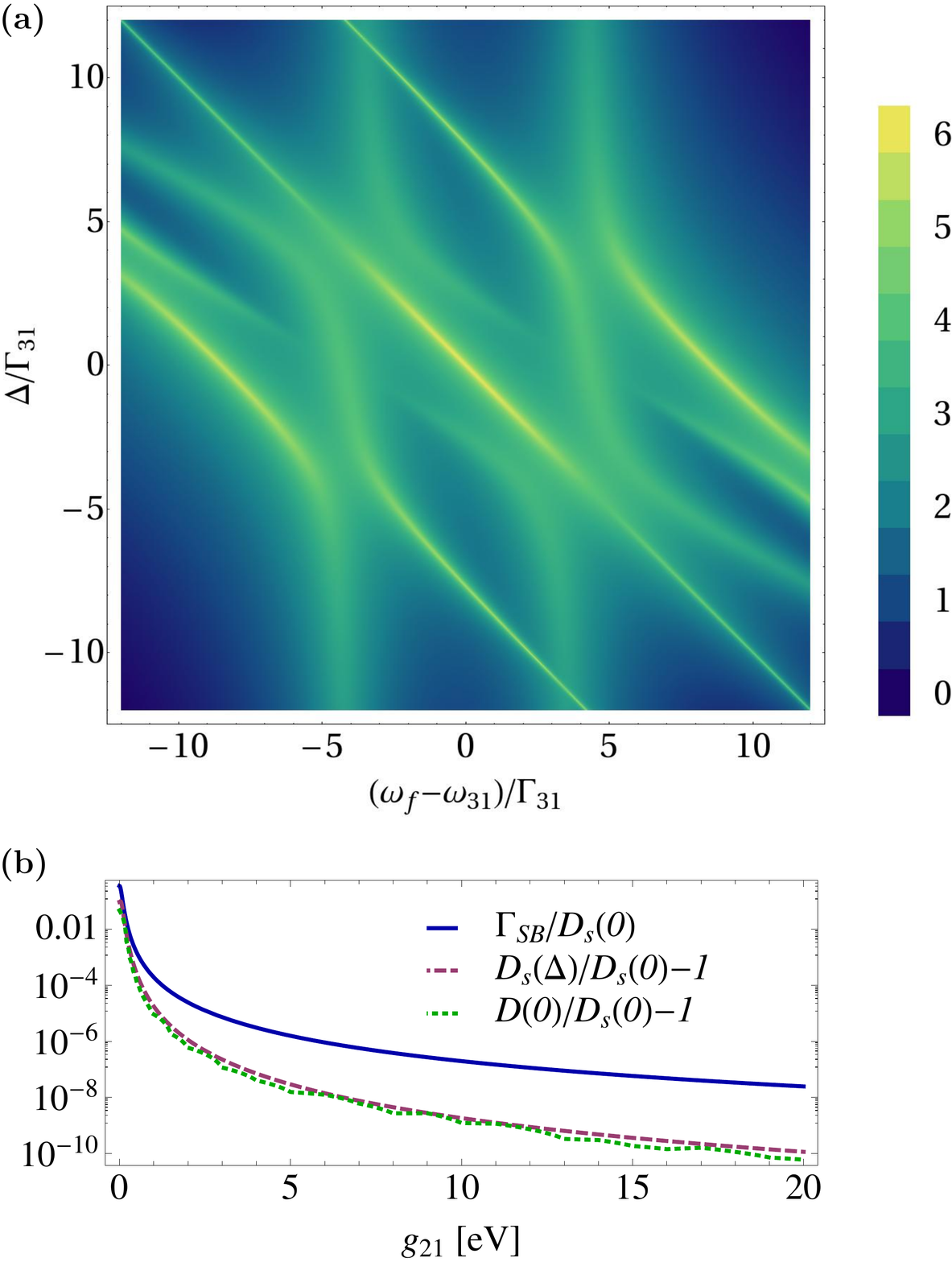}
\vspace{-5mm}
\caption{\label{FWHM}
(Color online.) 
(a) Density plot of the fluorescence spectrum (logarithmic scale, arb. units) as a function of the fluorescence
photon frequency $\omega_f$ with respect to the x-ray transition frequency $\omega_{31}$ (abscissa) and the laser detuning
$\Delta=\omega_x-\omega_{31}$ (ordinate), with the frequencies normalized by the $\Gamma_{31}$ rate.
The parameters are for Bi as in Fig.~\ref{fig:narrowing}.
(b) Continuous (blue) curve: ratio of the interference-narrowed width $\Gamma_{SB}$ of the outer sideband peaks to their
distance $D_s(0)=4G=4\sqrt{g_{31}^2+g_{21}^2}$ as a function of the optical Rabi frequency $g_{21}$,
with further parameters for the Bi three-level system as given in the third line of Table~\ref{thetable}.
Dashed (red) curve: deviation of the sideband distance $D_s(\Delta=\Gamma_{31})$ from $D_s(0)$.
Dotted (green) curve: deviation of the exact sideband distance $D(0)$ from its value in the secular limit $D_s(0)$.
}
\end{figure}

Transition lifetimes -- and related quantities like the atomic dipole or multipole moments --
are of great interest for astrophysical applications and for testing fundamental theories.
Measurements of these quantities are particularly necessary since they are
especially sensitive to the long-range behavior of atomic wave functions. Currently, even the best measurements
do not exceed the 10$^{-3}$ level of accuracy~\cite{LJC05short}. In our scheme, the narrowed central and outer lines enable
in principle an even more accurate determination of the atomic Rabi frequencies: the outer sideband peaks' distance
is given by $D_s(0)=4G=4\sqrt{g_{31}^2+g_{21}^2}$ (in the secular limit~\cite{remark} and when the x-ray laser is on
resonance, i.e. its detuning $\Delta=\omega_{31}-\omega_x$ from the transition frequency is 0).
In this formula, the optical Rabi frequency $g_{21}$ is usually known, therefore, determining its counterpart
$g_{31}$ for the x-ray transition is only limited by the accuracy of measuring the peak distance $D$.
Fig.~\ref{FWHM}~(b) shows the ratio of the width of the
narrowed outer lines to their distance $D_s(0)$. As shown, this ratio, characterizing the 
relative accuracy for the determination of atomic multipole moments, can be improved by several orders of magnitude
for higher optical laser intensities.

\color{black}{Our calculation, in analogy with the derivation of Ref.~\cite{Narducci1990} for zero detuning, shows that the outer sideband distance
for non-zero detuning is given by}\color{black}{}
$D_s(\Delta)=4G+\frac{G}{2}\left(4R-3R^2\right)(\Delta/G)^2+\mathcal{O}\left((\Delta/G)^4\right)$. 
This weak dependence is also illustrated in Fig.~\ref{FWHM}~(a). Hence, the experimental sensitivity on the
potentially inaccurately known detuning may be reduced by orders of magnitude by increasing the optical intensity
(Rabi frequency), as also shown in Fig.~\ref{FWHM} (b). The multipole matrix
elements of the ionic transitions can thus be determined in principle to a high accuracy on the order of $10^{-4}$--$10^{-6}$,
once the intensity of the driving field is accurately known. Conversely, the intensity may be measured to high
accuracy if the multipole moments are reliably known from an independent experiment (e.g. lifetime measurements).
At the same time, knowing the Rabi frequencies, the dependence of sideband positions on the x-ray detuning could
allow to measure in principle the ionic x-ray transition energy in an independent way.

\begin{table}[tb]
\begin{center}
\caption{
Parameters for $2s$$\leftrightarrow$$2p_{3/2}$ transitions in the Li-like ions ${}^{203}$Tl${}^{78+}$ ($\omega_{31}$=2236.5 eV),
${}^{209}$Bi${}^{80+}$ (2788.1 eV) and ${}^{235}$U${}^{89+}$ (4459.4 eV).
Optical transition energies
\color{black}{($\omega_{21}$,~\cite{Sha98BiHFS}),}\color{black}{}
natural line widths 
($\Gamma_{31}$, $\Gamma_{21}$) and Rabi frequencies ($g_{31}$, $g_{21}$) as well as the interference-narrowed
outer sideband width ($\Gamma_{SB}$) of the x-ray transition are given for the laser intensities $I_x$, $I_o$.
$x(y)$ stands for $x\times10^y$.
\label{thetable}
}
\begin{ruledtabular}
\begin{tabular}{ccccccccc}
               & $\omega_{21}$     & $\Gamma_{31}$ & $\Gamma_{SB}$ & $\Gamma_{21}$ &
$g_{31}$       & $g_{21}$       & $I_x$          & $I_o$       \\
\cline{2-7} \cline{8-9} 
               &  \multicolumn{6}{c}{[meV]} & \multicolumn{2}{c}{[W/cm$
               {}^{2}$]}         \\
\hline
Tl  & 499    & 6.6    & 7.1(-2) & 1.1(-12) & 1.8(2) & 2.1(3) & 1(12) & 1(16)  \\
~   & ~      & ~      & 7.2(-4) & ~        & 1.8(2) & 2.1(4) & 1(12) & 1(18)  \\
Bi  & 797    & 7.2(1) & 9.7(-2) & 7.7(-12) & 8.3(1) & 2.9(3) & 5(11) & 1(16)  \\
~   & ~      & ~      & 1.9(-1) & ~        & 1.2(3) & 2.9(4) & 1(14) & 1(18)  \\
U   & 136    & 2.4(1) & 3.7(-2) & 3.7(-14) & 7.7(1) & 2.8(3) & 5(11) & 1(16)  \\
~   & ~      & ~      & 1.3     & ~        & 3.3(4) & 1.9(5) & 9(16) & 5(19)  \\
\end{tabular}
\end{ruledtabular}
\vspace{-6mm}
\end{center}
\end{table}

Our above results have been demonstrated on the example of HCI with non-zero nuclear spins, i.e.
when hyperfine splitting of the ground state occurs. However, certainly, the results may be applied to further
three-level systems. For example, such configurations may also be prepared by applying (strong) external magnetic fields, which
gives rise to a large Zeeman splitting of the ground-state level, addressable by long-wavelength coherent radiation such as masers (or even
CO$_2$ lasers). Furthermore, the results can be generalized to other physical
systems with high transition energies, such as electromagnetic transitions in nuclei. In this setting, nuclear
multipole moments and transition energies may in principle be determined by an independent method.

Laser systems with photon energies of up to a few keV (in the range of Li-like 
transitions) are presently available~\cite{FLASH,LCLS}, allowing to excite elements as 
heavy as U and observing emission lines of sub-natural linewidths.
Upcoming laser facilities are expected to increase the frequency limit to the 
order of tens of keVs (e.g.~\cite{XFEL}), permitting to directly address the most relativistic heavy 
H-like systems or even nuclear transitions.
Table~\ref{thetable} lists values for some elements and atomic transitions.

In summary, a fully relativistic ab initio theory of the bound dynamics of atomic systems
in laser fields ranging to the x-ray domain has been developed.
The bare atomic states are constructed from solutions of the Dirac equation.
This approach allows for exploiting the sensitivity of inner-shell electrons to relativistic electron correlation,
QED and nuclear effects in strong Coulomb fields. As a demonstrative example, a means to determine
ionic transition multipole moments and frequencies via a three-level configuration 
driven by an x-ray and an optical field has been put forward. Current or 
near-future laser systems are expected to increase the accuracy of multipole moment determinations
from the current $10^{-3}$ level (via lifetime measurements) to the 10$^{-4}$ range or better. Furthermore,
the undesirable trapping of atomic population in a long-lived metastable state -- naturally
occurring in certain three-level systems -- can be reversed by the scheme presented here. Other
scenarios developed for the quantum control of non-relativistic resonance fluorescence
emission~\cite{Zhu96,Zhou96,Pasp98,Scully98,Keitel99,Kiffnershort,ScZu,Ficek} are anticipated to yield further improvement of detection and accuracy.

\begin{acknowledgments}

The authors acknowledge helpful conversations with Mihai Macovei and J{\"o}rg Evers.
Supported by Helmholtz Alliance HA216/EMMI.

\end{acknowledgments}


\begin{thebibliography}{27}
\bibitem[{rem()}]{remark}
\bibinfo{howpublished}{Our current discussion is valid in the secular limit
  $G=\sqrt{g_{31}^2+g_{21}^2} \gg {\rm
  max}\left\{\Gamma_{31},\Gamma_{32},\Gamma_{21}\right\}$. If the effective
  Rabi frequency $G$ can not be rendered high enough for a sufficient
  applicability of this approximation, one has to work with the more accurate
  and lengthy expressions of the spectral features.}
\end{thebibliography}
\end{document}